\documentclass[11pt,preprint]{aastex}

\shorttitle{NICMOS SURFACE PHOTOMETRY OF SEYFERT 2S}
\shortauthors{Martini, Pogge, Ravindranath, \& An}

\slugcomment{To appear in ApJ, November 2001}

\begin{document}

\newcommand{\um}{\mu{\rm m}}
\newcommand{\hst}{{\it HST\,\,}}


\title{
Hubble Space Telescope Observations of the CfA Seyfert 2s: Near-infrared 
Surface Photometry and Nuclear Bars\altaffilmark{1}}

\altaffiltext{1}{Based on observations with the
NASA/ESA {\it Hubble Space Telescope} obtained at the the Space Telescope
Science Institute, which is operated by the Association of Universities for
Research in Astronomy, Incorporated, under NASA contract NAS5-26555.}

\author{Paul Martini \altaffilmark{2}, Richard W. Pogge\altaffilmark{3}, 
Swara Ravindranath\altaffilmark{2,4}, Jin H. An\altaffilmark{3}}

\altaffiltext{2}{Carnegie Observatories, 813 Santa Barbara St., 
Pasadena, CA 91101-1292, martini@ociw.edu, swara@ociw.edu}

\altaffiltext{3}{Department of Astronomy, Ohio State University, 
140 W. 18th Ave., Columbus, OH 43210, pogge@astronomy.ohio-state.edu, 
jinhan@astronomy.ohio-state.edu}

\altaffiltext{4}{Department of Astronomy, University of California, 
Berkeley, CA 94720-3411}

\begin{abstract}

We present near-infrared (NIR) $J$ and $H$ surface photometry of 24 of the 
nearby Seyfert 1.8, 1.9 and 2 galaxies from the CfA Seyfert sample. 
The excellent angular resolution of the {\it Hubble Space Telescope} (\hst) 
probes spatial scales as small as 
tens of parsecs in most of these AGN and is sensitive to the 
presence of nuclear bars and other potential signatures of the AGN fueling 
process that channels host galaxy gas and dust to the nuclear region. 
We have used elliptical isophote techniques to search for nuclear bars in 
all of these galaxies and have employed a two-dimensional fitting technique 
to model the nuclear point source and surface brightness distribution of 
a bright subsample of these galaxies in an attempt to alleviate the impact 
of the nuclear point source on our sensitivity to nuclear bars. 
We find stellar nuclear bar candidates in four of these galaxies: 
Mrk~471, Mrk~270, Mrk~573, and NGC~5929, nearly 20\% of the total sample. 
The percentage rises to $\sim 30$\% when systems with disturbed morphologies 
or high inclinations are excluded. The nuclear bars in Mrk~573 and Mrk~270 
exhibit some evidence for dust lanes along their 
leading edges, analogous to the structures seen in host galaxy bars, while 
the dust lanes in Mrk~471 and NGC~5929 exhibit a more complex morphology. 
The fact that most of these AGN do not appear to contain stellar nuclear 
bars suggests that they are not the fueling mechanism for most 
low-luminosity AGN. 

\end{abstract}


\section{Introduction}

Bars and mergers have been the most commonly proposed fueling mechanisms for 
low-luminosity active galactic nuclei (AGN). While both of these methods 
can remove angular momentum from gas and dust, neither of these processes 
have been proven to be the sole arbiter of the AGN fueling process. 
Large-scale bars occur with equal frequency in normal and active galaxies 
\citep[][though see Knapen et al. 2000]{mcleod95,mulchaey97a,ho97a} and most 
low-luminosity AGN do not show any evidence of a recent major merger event 
\citep{fuentes88}. Recent, careful searches for evidence 
of minor mergers have also not found a clear excess of faint companions around 
these AGN \citep{derobertis98}. 

This lack of success in finding a definitive fueling mechanism for all 
low-luminosity AGN has driven observers to search at ever higher spatial 
resolution for signatures of the mechanisms that could remove sufficient 
angular momentum from gas and dust within the central few hundred parsecs. 
One small-scale feature that could fuel AGN activity 
\citep{shlosman89,pfenniger90} is a nuclear bar, which were first observed 
in nearby, large galaxies as enhancements in the surface brightness 
distribution \citep{devaucouleurs75,kormendy79,buta86a,buta86b}.
The nuclear bars first proposed to fuel AGN activity by \citet{shlosman89} 
were purely gaseous nuclear bars. In their model a large-scale bar leads to 
the transport of gas into the central region of the galaxy where it forms a 
circumnuclear gaseous disk. This disk could then become unstable to the 
formation of a purely gaseous, nuclear bar nested inside the larger bar 
and this gaseous nuclear bar could drive sufficient material inwards to 
fuel an AGN. Potentially the best method to detect gaseous nuclear bars 
is to observe their dust content in absorption against the background 
stellar light of the far side of the host galaxy. Gaseous nuclear bars may 
take the form of a bar-shaped dust lane and recently \citet{maiolino00} 
found a straight, or bar-shaped dust lane in Circinus (a Seyfert 2), where the 
gas kinematics were consistent with this interpretation.

Most of the nuclear bars observed to date have been stellar, rather than 
gaseous, nuclear bars as 
they correspond to clear enhancements in the visible or NIR surface 
brightness distribution. Stellar nuclear bars could also fuel AGN activity 
by removing angular momentum from circumnuclear material on small 
scales \citep{pfenniger90,friedli93}. As stellar nuclear bars appear to be 
randomly oriented with respect to host galaxy bars \citep{buta93}, fairly 
complex orbits are required to enhance the stellar density and create a 
stellar nuclear bar composed of old stars 
\citep{pfenniger90,maciejewski97,erwin99a,maciejewski00}. 
Stellar nuclear bars selected by their enhancement of the surface density 
have been found in several AGN hosts, including NGC~2681 and NGC~3945 
\citep{erwin99a}. A recent, kinematic study of four nuclear bar candidates 
selected on the basis of NIR surface brightness by \citet{emsellem01} found 
that the velocity fields of three of these galaxies were well-fit by a bar 
model. The fourth galaxy with a nuclear bar candidate instead hosts 
a kinematically decoupled gaseous disk and spiral structure within the 
inner Linblad resonance. 

Stellar nuclear bars could also form straight dust lanes similar to those 
observed in many strongly barred galaxies \citep[e.g. NGC~7479 
\& NGC~1530,][]{quillen95,regan97}. A strong triaxial potential can lead to 
significant gas inflow in large-scale bars \citep{athanassoula92} and 
if similar gas flow occurs in nuclear stellar bars, they may have 
associated dust lanes. \citet{regan99} argued that stellar nuclear bars could 
be more readily detected by searching for their influence on the ISM as 
the stellar surface density contrast may be quite weak due to the high 
velocity dispersion in galactic bulges.
They searched for such straight dust lanes in color maps of 12 Seyfert 
galaxies constructed from visible and NIR \hst images and found straight dust lanes extending into the nuclear 
region in NGC~3081, NGC~5347, and NGC~7743. 
\citet{martini99} performed a similar search with $V-H$ and $J-H$ \hst color 
maps and found straight dust lanes in Mrk~573, Mrk~270, and Mrk~471 in 
the sample of 24 Seyfert 2s we describe in this paper. Recently, 
\citet{maciejewski01} have argued that in dynamically stable 
configurations the nuclear bar does not extend to its corotation radius and 
thus it will not form the strong shocks and corresponding dust 
lanes seen in large-scale bars sought by \citet{regan99} and \citet{martini99}. 

Both of these studies of relatively large numbers of AGN concluded that 
stellar nuclear bars were present in only a minority of AGN based on 
selection by the presence of straight dust lanes, rather than enhancements in 
the stellar surface density. \citet{martini99} also searched by visual 
inspection for stellar nuclear bars in the NIR surface brightness contours 
and found most of the galaxies with straight dust lanes also 
showed evidence for nuclear bars in the stellar distribution, yet they did 
not employ a quantitative set of selection criteria. In addition, the results 
of \citet{maciejewski01} suggest that many nuclear bars would be missed 
in selection based on the presence of straight dust lanes. 
For the remainder of 
this paper we will concentrate on stellar nuclear bars selected by their 
associated starlight and refer to them as simply nuclear bars; we will  
explicity refer to stellar nuclear bars selected by dust 
morphology or purely gaseous nuclear bars when appropriate. 
Nuclear bars are also commonly referred to as secondary bars when they 
are found in galaxies with large-scale bars, although we will not use this 
terminology as our sample was not previously selected to only contain galaxies 
with large-scale bars. 

The recent visible and near-infrared (NIR) studies of the circumnuclear 
regions of low-luminosity AGN with \hst described above also found that most 
contain dusty ``nuclear'' spiral structure on 100s of parsec scales that is 
distinct from the main disk spiral arms. These studies found that many AGN 
have dusty nuclear spiral structure \citep{quillen99,regan99,martini99}.
Similar nuclear spiral structure has also been observed in a number of 
quiescent galaxies \citep{phillips96,carollo98a,elmegreen98,laine99}. 
\citet{martini99} estimated that the gaseous disks 
were not self-gravitating based on a measure of the total extinction in the 
nuclear spiral arms. They concluded that these spiral structures could be 
due to shocks propagating in circumnuclear gaseous disks, a mechanism 
proposed by \citet{elmegreen98} and \citet{montenegro99}. Such shocks could 
dissipate sufficient angular momentum to fuel these low-luminosity AGN, 
which only require mass accretion rates of $\sim 0.01 - 0.1 M_\odot$ yr$^{-1}$. 
Nuclear spiral structure is a tempting signature of the AGN fueling process 
because, unlike nuclear bars and interactions, it appears to be present in 
all AGN that do not have morphologically disturbed circumnuclear regions. 
While there has not yet been a systematic control study to assess the relative 
frequency of nuclear spiral structure in a control sample of quiescent 
galaxies, recent \hst studies of nearby spirals have not found that nuclear 
spiral structure is similarly ubiquitous in normal galaxies 
\citep[e.g.][]{carollo98a}. 

In this paper we present a detailed study of the NIR surface photometry 
from \citet{martini99}, which include 24 of the 25 Seyferts classified as 
type 1.8, 1.9, or 2 by \citet{osterbrock93} from the CfA Redshift 
Survey \citep{huchra92} and listed in Table~\ref{tbl:sample}. 
The main goal of the present paper is to reanalyze the nuclear bar fraction in 
this sample using a quantitative bar detection algorithm based on the NIR 
surface brightness distribution in order to test the hypothesis that nuclear 
bars are responsible for fueling all low-luminosity AGN activity. 
Given the recent theoretical investigation of \citet{maciejewski01}, 
dynamically stable nuclear bars may not form the straight dust lanes used 
to select previous nuclear bar samples and therefore nuclear bars selected 
by the NIR surface brightness distribution may provide a stronger constraint 
on the importance of nuclear bars to the AGN fueling process. 
In addition to our search for nuclear bars, 
we also measure the surface brightness profiles for all of these galaxies and 
derive the best-fit profile parameters and nuclear luminosities for a subset 
of galaxies that are sufficiently bright and not in edge-on or morphologically 
disturbed systems. 

\section{Image Processing} \label{sec:proc}

These images were all obtained with the NICMOS Camera 1 (NIC1) on 
\hst, which has a plate scale of $0.043''$ pixel$^{-1}$. With the exception 
of NGC~1068 \citep{thompson99}, all of these galaxies were observed as 
part of GO-7867. We observed them through the F110W and F160W filters for 
1024s per filter, split into four dither positions (SPIRAL-DITH) 
of 256s and offset one arcsecond to aid bad pixel 
rejection.  The F110W and F160W filters (F110M and F170M for NGC~1068) are 
roughly equivalent to the ground-based $J$ and $H$ filters, respectively. 

The images were processed through the standard CALNICA data reduction 
pipeline from STScI. As noted by \citet{martini99}, we did not use the CALNICB 
part of the STScI pipeline to combine our four dithered images. 
CALNICB attempts to subtract the sky level from an image sequence when 
it shifts and adds the individual dither positions. In our images, however, 
even the faintest objects fill nearly half of the field, and so a sky 
subtraction is impossible. In any event, the sky background for these images 
through the $J$ and $H$ filters is negligible based on measurements of both 
the more compact galaxies in our sample and archival, blank fields with the 
same instrument configuration. Instead of CALNICB, we used simple integer 
shifts to align and stack the four dither positions. 

The photometric calibration of \hst data is in general 
nontrivial due to differences between \hst filters and their nearest 
ground-based analogs. The F110W and F160W filters in NICMOS are broader than 
the standard, ground-based $J$ and $H$ filters and include spectral regions 
in these galaxies that are unobservable from the ground due to telluric water 
absorption. To transform our images to the ground-based CTIO/CIT 
photometric system \citep{elias82,persson98}, we used the photometric 
zeropoints established by the NICMOS photometric calibration program and the 
color terms calculated by \citet{stephens00}. \citet{stephens00} calculated 
a complete photometric solution for NICMOS Camera 2 (NIC2) based on 
observations of red standards, in addition to the bluer standards observed as 
part of the NICMOS photometric calibration program. To calibrate our NIC1 data, 
we used the STScI zeropoint calibration and supplemented it with the color 
terms from \citet{stephens00}. While these color terms were derived for a 
different camera, they should be dominated by the wavelength dependence of 
the filter transformation and array quantum efficiency and these quantities 
are nearly identical for NIC1 and NIC2. For NGC~1068 we used the standard 
NICMOS photometric calibration. 
 
\section{Surface Photometry and Nuclear Bars} \label{sec:surf}

\begin{figure} 
\epsscale{.8}
\plotone{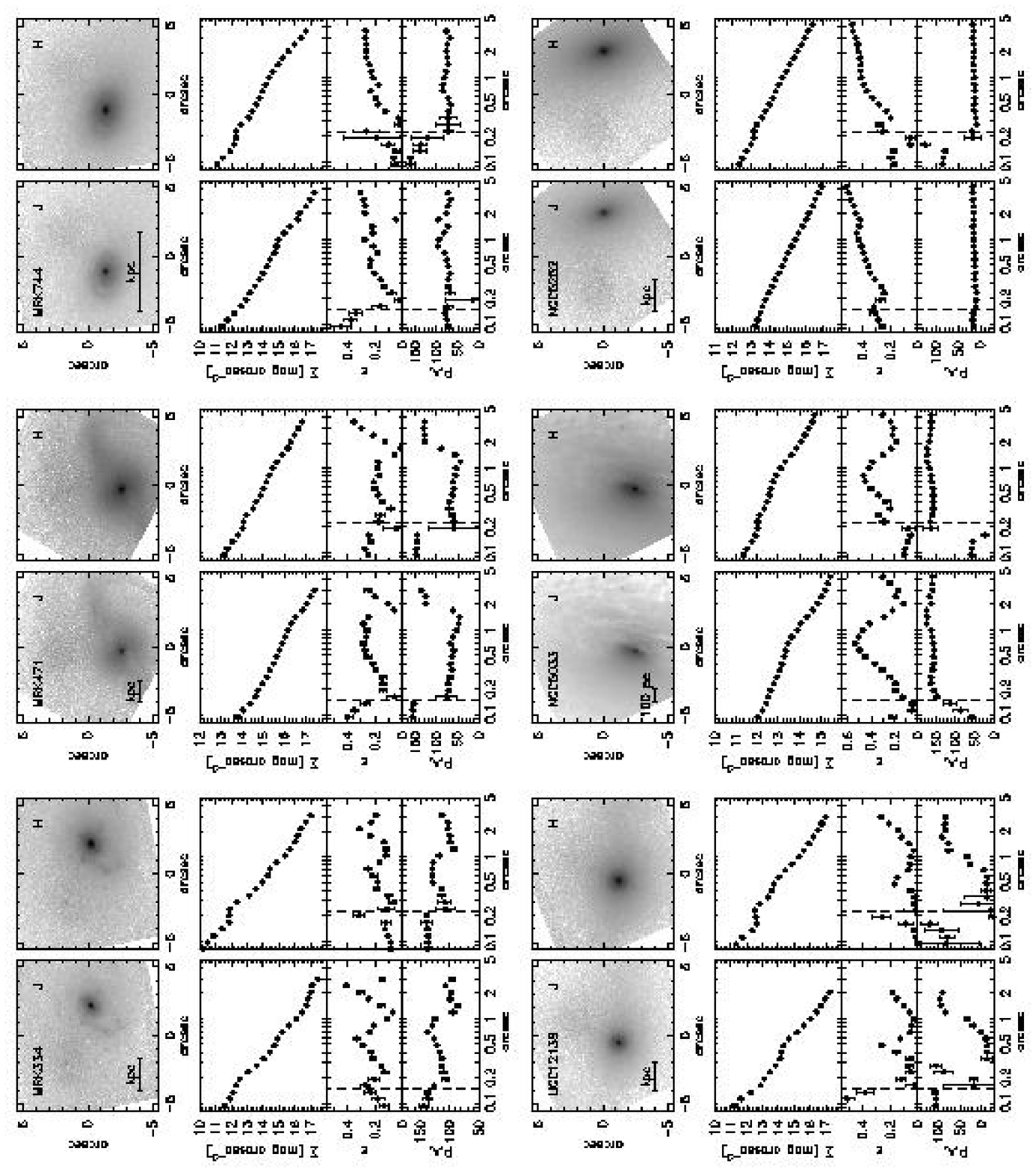}
\caption{Grayscale images and surface brightness, ellipticity, and position 
angle profiles as a function of semimajor axis for Mrk~334, Mrk~471, Mrk~744, 
UGC~12138, NGC~5033, and NGC~5252.  The left panels in each plot correspond 
to the $J$ image of the galaxy and the right panels to the $H$ image. 
The grayscale images ({\it top panels}) are on a log scale and have been 
rotated so that north is up and east is to the left. The remaining panels show 
the surface brightness profile, ellipticity, and position angle (measured 
north through east) as a function of the ellipse semimajor axis. The dashed, 
vertical lines mark the location of first Airy ring, and designate the 
smallest angular scale at which we could detect nuclear bars (see 
Section~\ref{sec:surf}). 
\label{fig1} }
\end{figure}

Figures~\ref{fig1} -- \ref{fig4} show $J$ and $H$ images 
({\it top panels}) for this sample. These grayscale
images show the wealth of 
different morphologies present in the central regions of Seyferts. 
One qualitative trend is the strength of the nuclear point source 
with Seyfert type. The Seyfert~2 galaxies have on average weaker nuclear 
point sources than the 1.9s and 1.8s. This agrees with the trend 
observed in previous visible-wavelength \hst observations of Seyferts, 
where Seyfert~1s tend to have extremely bright nuclear sources and Seyfert~2s 
have a significantly weaker nuclear contribution \citep{nelson96,malkan98}. 
Two interesting exceptions are the extremely strong nuclear point sources 
in the Seyfert~2s NGC~1068 and NGC~7674 (Figures~\ref{fig3} and \ref{fig4}), 
which harbor polarized broad line regions \citep{miller83,miller90}. To a 
lesser extent Mrk~573 and NGC~5347 (Figures~\ref{fig3} and \ref{fig4}) 
appear to have stronger nuclear point sources than the other Seyfert~2s. 

We used the ELLIPSE task in IRAF\footnote{IRAF is distributed by the National 
Optical Astronomy Observatories, which are operated by the Association of 
Universities for Research in Astronomy, Inc., under cooperative agreement with 
the National Science Foundation.} STSDAS to fit elliptical isophotes to these 
galaxies and measure the surface brightness, ellipticity ($\epsilon = 
1 - b/a$), and position angle of the semimajor axis (PA, in degrees 
measured north through east). ELLIPSE fits elliptical isophotes according to 
the formalism outlined by \citet{jedrzejewski87} and includes measurement of 
higher order Fourier coefficients that characterize the deviation of the 
isophotes from perfect ellipses. The lower panels in Figure~\ref{fig1} -- 
\ref{fig4} ({\it top to bottom}) show 
the $J$ and $H$ surface brightness profile, ellipticity, and position angle 
as a function of semimajor axis. The quality of the flat field and instrumental 
background subtraction can affect the elliptical isophotes on large scales. 
The nuclei of nearly all of these galaxies fall on one of the lower two 
quadrants of the NIC1 array, which have higher quantum efficiency than the 
upper-left quadrant. The low quantum efficiency of the upper-left quadrant 
requires a large flatfield correction and uncertainties in the bias level and 
the flatfield could systematically distort the elliptical isophote fits. 
In order to verify that we were not affected by this systematic source of 
error, we constructed a data quality file for the ELLIPSE package that masked 
out the lowest quantum efficiency regions in the upper-left quadrant of the 
NIC1 array. We also visually inspected the ellipse fits to each galaxy 
to insure that the position angle at large semimajor axis was not artificially 
twisted by any errors in the flatfield. 

On smaller angular scales, the structure in the NICMOS point spread 
function (PSF) can significantly affect the quality of the fits to some of 
these galaxies. This is particularly 
striking in the Seyfert~1.8 galaxies (e.g. Mrk~334 or UGC~12138, 
Figure~\ref{fig1}) where 
the ``bump'' clearly visible in the surface brightness profile is due to the 
first Airy ring of the PSF at $r_J \sim 0.15''$ and $r_H \sim 0.23''$. 
The ellipse fits for strong nuclear point sources also tend to include a 
strong fourth-order Fourier coefficient due to the nonaxisymmetric, 
``boxy'' diffraction pattern in the PSF. These cause the 
ellipticity and position angle distributions to have large errors and 
scatter until the influence of the first Airy ring falls off at 
$\sim 0.3''$, or even until higher order diffraction patterns diminish at 
$\sim 0.7''$, such as in Mrk~334, UGC~12138, or NGC~1068.
For galaxies with little or no nuclear contribution, the ellipticity 
and position angle distributions are reliable for $r_J \geq 0.15''$ and 
$r_H \geq 0.23''$. 
The vertical, dashed lines in Figures~\ref{fig1} -- \ref{fig4}
correspond to these lower limits in the sensitivity of the measured 
ellipticity and position angle distributions to nuclear bar candidates. 

To quantify the 
sensitivity of our NIR surface photometry to the presence of nuclear bars, 
we adopt the criteria used by \citet{mulchaey97b} to identify 
host galaxy bars: an increase in ellipticity at constant position angle 
followed by a drop in ellipticity to the inclination of the disk. As the 
$11''$x$11''$ NIC1 FOV does not always include much of the host galaxy disk, 
the final position angle and ellipticity may reflect the presence of a host 
galaxy bar; we therefore relax the latter criterion. By examination of the 
ellipticity and position angle profiles in the figures, 
we find nuclear bar candidates in Mrk~471, Mrk~573, Mrk~270, and NGC~5929. 
Figure~\ref{fig5} shows the color maps from \citet{martini99} along with the 
$H-$band surface brightness contours for these four galaxies. 
All of these galaxies were suggested to have nuclear bars by \citet{martini99} 
based on the appearance of the nuclear surface brightness isophotes, and 
all but NGC~5929 exhibit straight dust lanes in their visible-NIR color maps. 
NGC~5347, which was suggested by \citet{regan99} to have the straight dust 
lanes indicative of a nuclear bar, does not show strong evidence of a nuclear 
bar in the NIR surface brightness distribution. The lack of NIR 
morphological evidence precludes the presence of a significant old stellar 
population associated with a nuclear bar. However, as demonstrated by 
\citet{regan99}, a small contrast in the stellar density may excite an
order of magnitude larger contrast in the ISM.

\citet{knapen00} define a galaxy as barred if the ellipticity varies by 
$\epsilon \geq 0.1$ over a region of constant position angle or if the 
position angle changes by $\geq 75$\degr\, over a range where the 
ellipticity is greater than $0.1$. All four of the galaxies 
classified as barred according to the \citet{mulchaey97b} classification 
scheme would also be barred under these similar criteria. Mrk~573 and 
Mrk~270 meet the first set of criteria as they change in ellipticity at 
constant position angle, while the remaining two galaxies have changes in 
position angle corresponding to their ellipticity variations. 

The properties of the nuclear bar candidates are summarized in 
Table~\ref{tbl:bars} and a description of the results for individual 
galaxies are given in the Appendix. Both Mrk~573 and Mrk~471 have 
prominent, large-scale bars and therefore could also be termed 
secondary bars. The remaining two nuclear bars candidates, Mrk~270 and 
NGC~5929, do not have known large-scale bars. Given their small-scale, 
they are clearly nuclear bars, but they are not also secondary 
bars. Mrk~270 was one of the few CfA Seyferts not observed by \citet{mcleod95} 
in their $K-$band survey, and it is only typed as "SO?" in the RC3 catalog. 
NGC~5929 is classified as unbarred in the RC3 and by \citet{mcleod95}. 
However, this galaxy is in the midst of an interaction which could have 
obscured or destroyed the large-scale bar. Another possibility is that the 
interaction may have sufficiently disturbed the small-scale morphology 
to artificially produce the nuclear bar candidate in the NIR isophotes. 

As discussed above, the complex NICMOS PSF can introduce significant 
scatter into the ellipticity and position angle distributions used to identify 
nuclear bar candidates. Our sensitivity to candidates of a given physical 
length is therefore sensitive to the distance of a galaxy and the strength 
of its nuclear point source. This issue is discussed in 
section~\ref{sec:discuss}. For galaxies with sufficient signal-to-noise we 
have constructed two-dimensional models of the galaxy surface brightness 
distribution and nuclear point source in an attempt to improve our sensitivity 
to nuclear bars and parameterize the central light distribution of these 
Seyferts. This technique is described in the next section.  

\section{Analysis} \label{sec:analysis}

The high angular resolution of \hst that has made it possible to systematically 
search for nuclear bars and spiral structure in nearby AGN has also 
revolutionized the study of the central stellar light distribution in galaxies
\citep[e.g.][]{lauer95,faber97}. Nearly all early-type galaxies with dust 
in their central regions have compact nuclear sources 
\citep{lauer95,vandokkum95,rest01,ravindranath01}. While all later-type 
galaxies usually have significant dust in their nuclear regions, irrespective 
of whether or not they host an AGN or nuclear star formation, 
\citet{carollo98b} suggested that compact nuclear sources are more 
common in galaxies with exponential rather than $R^{1/4}$ bulges and that 
exponential bulges likely have lower stellar densities.  \citet{marquez99} 
explored the circumnuclear properties of isolated spirals with and without 
AGN based on ground-based NIR observations and found no differences in the 
profile shapes of their AGN host and normal galaxy samples, although they 
did find that the central colors of AGN are redder than normal spirals 
\citep{marquez00}. 

A two-dimensional analysis of the surface brightness distribution 
was performed in order to quantify the contributions from the galaxy 
bulge and the nuclear point source. The 2-D decomposition of the
components was done using the least--squares fitting program 
GALFIT \citep{peng01}, which models the galaxy light with a combination 
of various analytic functions (e.g. S\'ersic, de Vaucouleurs, Nuker, 
exponential, Gaussian, Moffat).  GALFIT can also simultaneously fit an 
additional point source (AGN or compact star cluster) and provide a good 
measurement of the nuclear magnitude.

One of the main concerns while trying to obtain information at the 
highest spatial resolution is to account for the effects of the PSF. 
Our images do not have the high signal-to-noise ratios 
required for using deconvolution techniques. Instead, the 2-D modeling 
routine used here convolves the galaxy model with high-quality synthetic 
PSFs produced by the TINYTIM software \citep{krist99}. The TINYTIM 
PSFs are adequate for these applications even though there may be 
uncertainties due to temporal variations like thermal changes in the 
instrument and shifts in the pupil mask alignment \citep{krist97}. 
  
We fit the galaxy bulge using either a ``Nuker'' law \citep{lauer95}, 
which has the form: 
\begin{equation}
I(r) = 2^{(\beta-\gamma)/\alpha} I_b \left(\frac{r}{r_b}\right)^{-\gamma} \left[1 + \left(\frac{r}{r_b}\right)^\alpha \right]^{(\gamma-\beta)/\alpha}, 
\end{equation}
or a S\'ersic profile \citep{sersic68} of the form:
\begin{equation}
I(r) = I_e {\rm exp} \left( -b \left[ \left(\frac{r}{r_s}\right)^{1/n} -1 \right] \right)
\end{equation}
and the nucleus was modeled with a Gaussian function. The details of the
fitting procedure are described by \citet{ravindranath01}. 

A major difficulty in trying to decouple the galaxy contribution from the point 
source arises from the relatively small field of view of the NIC1 images. In 
a few cases the nucleus is very bright and the diffraction rings are
prominent out to $1''$ semi-major axis, leaving only a small region for
sampling the contribution from the galaxy (e.g., NGC~5674 and NGC~7674, 
see Figures~\ref{fig2} and \ref{fig4}). We could obtain reasonably good 
2-D fits for eleven galaxies in the sample. In most cases, we were able to 
parametrize the surface brightness using a Nuker function, although for 
two galaxies we were only able to fit S\'ersic profile as it has fewer 
parameters. The inner slope ($\gamma$) values for all the galaxies fall in 
the range 0.45-0.7 and imply steeper central surface brightness profiles than 
seen in many earlier-type galaxies \citep{ravindranath01}. 

The nuclear apparent magnitudes of these AGN are a measure of the luminosity 
of the accretion onto the central black hole. Combined with observations at 
other wavelengths, these measurements probe the spectral energy distribution 
for the accretion process \citep[e.g.][]{ho99}. The apparent brightness of 
many of these nuclear point sources were previously measured by 
\citet{quillen01} and the measurements 
are in good agreement within the uncertainties. The best-fit galaxy profile 
parameters along with the apparent magnitude for the nuclear point source are 
given in Table~\ref{tbl:profs}. 

The 2-D models are generated for a fixed ellipticity and position 
angle determined from the isophotal contours in the outer region. 
Therefore the residual image formed from the difference of the galaxy and 
model fit enables us to identify features (e.g. dust lanes and bars) 
that cause significant changes in the ellipticity and position angle.
Since the 2-D analysis includes a fit to the nuclear point source as an
additional component, the fit residuals provide a more sensitive probe
of underlying nuclear bars at small semi-major axis length in galaxies with 
a bright nucleus than the simple elliptical isophote fits discussed in
section~\ref{sec:surf}. The nuclear bars in Mrk~270, Mrk~573 and NGC~5929 are 
clearly visible in the residual images, but no additional 
nuclear bar candidates were recovered with this technique. 

The remaining galaxies could not be fit with 2-D models due to either 
insufficient signal-to-noise in the galaxy component or 
irregular/peculiar morphology.  For example, Mrk~334 or 
NGC~4388 are obviously too irregular or dusty to fit with smooth elliptical 
isophotes. More regular galaxies such as NGC~3362 or UM~146 are sufficiently 
faint that the galaxy is undetected over much of even the small NIC1 
field-of-view. Finally, while Seyferts like NGC~5347 and UGC~12138 are 
bright, the nuclear PSF dominates the signal from the galaxy out to nearly 
$1''$ and at larger radii the galaxy is too faint for an acceptable fit. As 
it is not possible to accurately model the galaxy light distribution for these 
remaining objects in our sample, we cannot reliably measure the apparent 
magnitude of the nuclear point source. 

\section{Discussion} \label{sec:discuss}

There are now a large number of galaxies with nuclear bars \citep{buta93,
friedli96,jungwiert97} and no evidence for any preferred position 
angle of the nuclear bar with respect to the host galaxy bar. The apparently 
random orientation between the host and nuclear bars is consistent with 
models that do not predict the nuclear and host galaxy bar to be kinematically 
coupled \citep[e.g.][]{friedli99}. 

Of the 24 galaxies in our sample, five (Mrk~266, Mrk~334, Mrk~744, NGC~4388, 
and NGC~5033) are sufficiently disturbed or high inclination systems in 
which we would not expect to see a nuclear bar if one were present. In the 
remaining 19 galaxies, our sensitivity to nuclear bars of a given semimajor 
axis length is a function of distance, galaxy inclination, and the brightness 
of the nuclear point source. The {\it gaseous} nuclear bar in Circinus studied 
by \citet{maiolino00} has a semimajor axis length of approximately 100pc 
and is one of the shortest 
known nuclear bars.  We would have detected nuclear bars 100pc or longer 
in 13 of these galaxies and in fact do detect a bar in four, 
with projected semimajor axes ranging in size from nearly 300 to 900pc 
(see Table~\ref{tbl:bars}). 
For the remaining six, more distant galaxies, the average minimum projected 
semimajor axis length we are sensitive to is $\sim 160$pc. While in many 
cases the NIC1 field of 
view is not large enough to be sensitive to nuclear bars that extend to a 
kiloparsec, ground-based NIR imaging of these galaxies by \citet{mcleod95} and 
\citet{mulchaey97b} were sensitive to nuclear bars at these scales. 

All four of our nuclear bar candidates meet the bar selection criteria used 
by \citet{mulchaey97b} and \citet{knapen00} to select host galaxy bars. 
Mrk~573 and Mrk~270 have dust lanes along one edge of the nuclear bar in 
the visible--NIR colormaps (see Figure~\ref{fig5}) which then turn sharply 
inwards to the nucleus, forming the straight dust lanes across the nucleus 
observed by \citet{martini99}. This dust morphology is similar to what is 
observed in host galaxy bars \citep{quillen95,regan97}. If these dust lanes 
do trace shocks along the leading edge of the nuclear bar, they must extend 
to their corotation radius to produce such strong shocks 
\citep{maciejewski01}.  
The dust morphology associated with the nuclear bars in Mrk~471 and NGC~5929 
appears more irregular. Mrk~471 does also show evidence for a straight dust 
lane crossing the nucleus perpendicular to the nuclear bar position angle, 
similar to what is seen in Mrk~573 and Mrk~270, but there is no evidence 
for dust along the nuclear bar at larger radii. The circumnuclear region of 
NGC~5929 is quite irregular. There is some clumpy dust perpendicular to the 
nuclear bar position angle, but no evidence for coherent structures on 
larger scales that are associated with the nuclear bar. 

Because the nuclear bars discovered in nearby galaxies have small angular 
sizes, only a small fraction of the $\sim 40$ known nuclear bars 
\citep{friedli99} have kinematic data available. \citet{maiolino00} obtained 
kinematic information for the gaseous nuclear bar in Circinus that supported 
their nuclear bar interpretation, while \citet{emsellem01} obtained spectra 
of four nuclear bars in nearby galaxies and found that three of them were 
well-fit by a nuclear bar model. These studies suggest that there are true 
nuclear bars and, from the study of \citet{emsellem01}, that they can be made 
up of old stars. Kinematic information for the four nuclear bar candidates 
in these Seyferts could confirm that they follow the dynamics expected for 
nuclear bars. 

We have found nuclear bar candidates in four of the 24 Seyferts in this 
sample using a well-defined, quantitative selection technique based on the 
NIR surface brightness distribution. 
Our observations were sensitive to nuclear bars with a semimajor axis length 
as small as $\sim 100$pc for 13 of these galaxies and to approximately twice 
this length for six additional galaxies. Previous studies of the frequency of 
nuclear bars in spirals generally found them in 20 -- 30\% of galaxies 
\citep{buta93,friedli96,regan99,erwin99b,marquez00}, in good agreement with 
the fraction we report here, although the selection techniques and spatial 
resolution vary considerably between these different investigations. Our 
result that nuclear bars are present in only a minority of AGN strongly 
suggests that they are not responsible for removing angular momentum and 
transporting fuel from the host galaxy to the nuclear ($\sim 10$ pc) region in 
most Seyferts. As we have used a well-defined and quantitative method of 
selecting nuclear bars, the relative frequency of nuclear bars in AGN and 
non-active galaxies could be studied with a future control sample.  

\acknowledgments

We would like to thank Chien Peng for providing us with a copy of the 
GALFIT code and both referees for some helpful comments. Witold 
Maciejewski also provided us with many useful comments that helped to 
clarify the discussion of the different types of nuclear bars. 
Support for this work was provided by NASA through grant number
GO-07867.01-96A from the Space Telescope Science Institute, which is 
operated by the Association of Universities for Research in Astronomy, 
Inc., under NASA contract NAS5-26555.  This research has made use of the
NASA/IPAC Extragalactic Database (NED) which is operated by the Jet Propulsion
Laboratory, California Institute of Technology, under contract with the
National Aeronautics and Space Administration.

\appendix

\section{Notes on Individual Objects}

For each galaxy we briefly summarize the NIR morphology and discuss the 
quality of the elliptical isophote and 2-D fits. We also mention any 
previous observations which may affect our interpretation. 

\noindent
Mrk 334   -- This interacting system has a tidal arm visible in both the 
$J$ and $H$ images. The nuclear point source is particularly 
bright: both the first Airy ring and a second diffraction ring is visible 
at $0.7''$ in the $H$ image. The presence of 
the PSF features results in limited sensitivity to any potential nuclear bar 
with semimajor axis $< 310$ pc. Because this is an interacting system, a 
nuclear bar is probably not required to remove angular momentum from gas 
in the host galaxy to transport it inwards. 

\noindent
Mrk 471   -- The nuclear bar candidate in this galaxy is apparent in 
grayscale images, as is the break in the ellipticity and 
position angle at $\sim 1.3''$, which corresponds to a projected $\sim 860$ 
pc semimajor axis length. This projected semimajor axis length is comparable 
to some host galaxy bars (e.g. NGC~1068), yet this galaxy also has a clear 
host galaxy bar, which is apparent on larger scales in the grayscale. 
The straight dust lanes noticed by \citet{martini99} are 
nearly perpendicular to the position angle of the nuclear bar within the 
bar semiminor axis, although they do not appear to trace the edge of the bar 
at larger radii, as seen in Mrk~270.

\noindent
Mrk 744   -- \citet{keel80} first noticed that Seyfert 1s are found much less 
frequently in edge-on systems compared Seyfert 2s. Dust obscuration from the 
host galaxy was suggested to be particularly important in this high-inclination 
system by \citet{goodrich83} and the bright, NIR nuclear point source is 
further evidence that it only appears as a weak broad-line system due to host 
galaxy dust. The elliptical isophotes only poorly fit the morphology of 
this inclined and interacting system.

\noindent
UGC 12138 -- Interpretation of the circumnuclear structure in this galaxy 
is complicated by the strong nuclear point source. As in Mrk~334, the first 
Airy ring and a second ring at $0.7''$ are visible in the $H$ surface 
brightness profile. These features limit the sensitivity to 
a nuclear bar with semimajor axis $< 350$ pc. At larger radii, the 
ellipticity starts to increase again at the position angle of the host galaxy 
bar.

\noindent
NGC 5033  --  This nearly edge-on galaxy hosts a Seyfert 1.9 nucleus and 
is conspicuously brighter in the NIR than at visible wavelengths, suggesting 
that host galaxy dust contributes to the obscuration of the 
nuclear region. The position angle of $\sim 160$ degrees measured in the (poor) elliptical isophote fits is in good agreement with the value measured by 
\citet{thean97}. 

\noindent
NGC 5252  -- The nuclear point source in this Seyfert 1.9 is relatively weak 
and only slightly affects the ellipticity distribution for $< 0.2''$. This 
galaxy is well-fit by a Nuker profile.

\noindent
NGC 5273  -- There is a ``spike'' in the ellipticity distribution 
at $r_J \sim 0.5''$ or $r_H \sim 0.6''$ (corresponding to a projected 
semimajor axis length of $\sim 40$ pc). While this spike is of sufficient 
amplitude to meet our nuclear bar selection criteria, the fact that the 
variation in ellipticity occurs at a larger semimajor axis in the longer 
wavelength 
filter suggests that this is probably an artifact of the strong nuclear point 
source. Dust may also be a factor as the ellipticity variation is greater at 
$J$ than at $H$. For these reasons we do not classify this spike as a 
nuclear bar candidate. 

\noindent
NGC 5674  -- While there is a large ellipticity variation at $J$, it is not 
present at $H$. There is a significant amount of amorphous dust perpendicular 
to the position angle of this ellipticity, thus the ellipticity change is 
very likely due to dust. The nuclear point source is also bright in this 
galaxy and it clearly compromises the elliptical isophote fits to the inner 
region. We therefore do not consider this to be a nuclear bar candidate. 
The underlying galaxy was otherwise sufficiently bright and regular for 
GALFIT to model the surface brightness distribution. 

\noindent
UM 146    -- There is a clear jump in the $H$ surface brightness profile 
due to the first Airy ring of the nuclear PSF. The galaxy appears otherwise 
quite faint and compact; the elliptical isophotes only extend to $\sim 2''$. 

\noindent
Mrk 461   -- This galaxy has a relatively smooth surface brightness profile, 
but there was not sufficient signal-to-noise over a large enough range in 
radius to fit the underlying galaxy profile shape. 

\begin{figure} 
\epsscale{1.}
\plotone{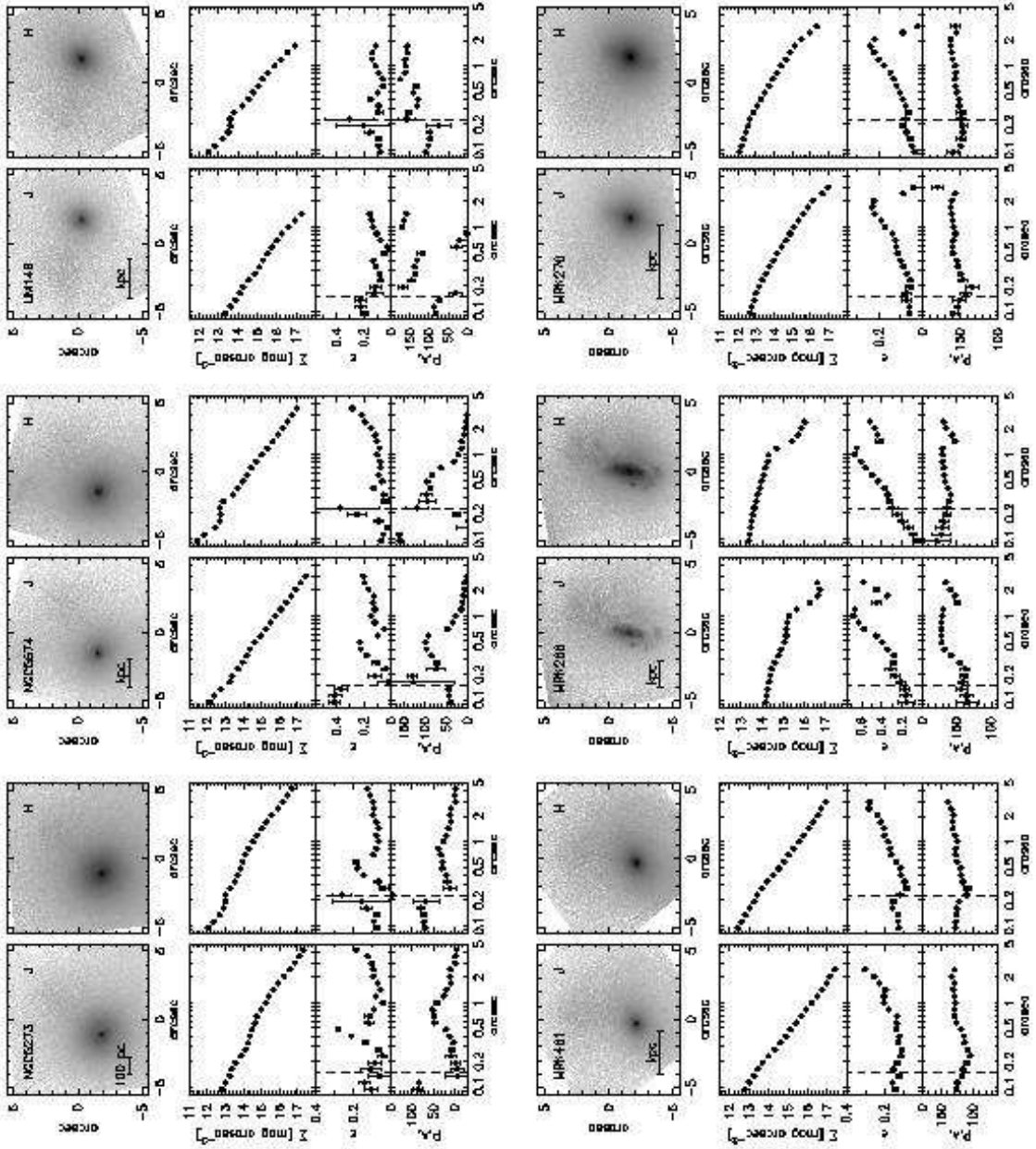}
\caption{Same as Figure~\ref{fig1} for NGC~5273, NGC~5674, UM~146, 
Mrk~461, Mrk~266SW, and Mrk~270. \label{fig2} }
\end{figure}

\begin{figure} 
\epsscale{1.}
\plotone{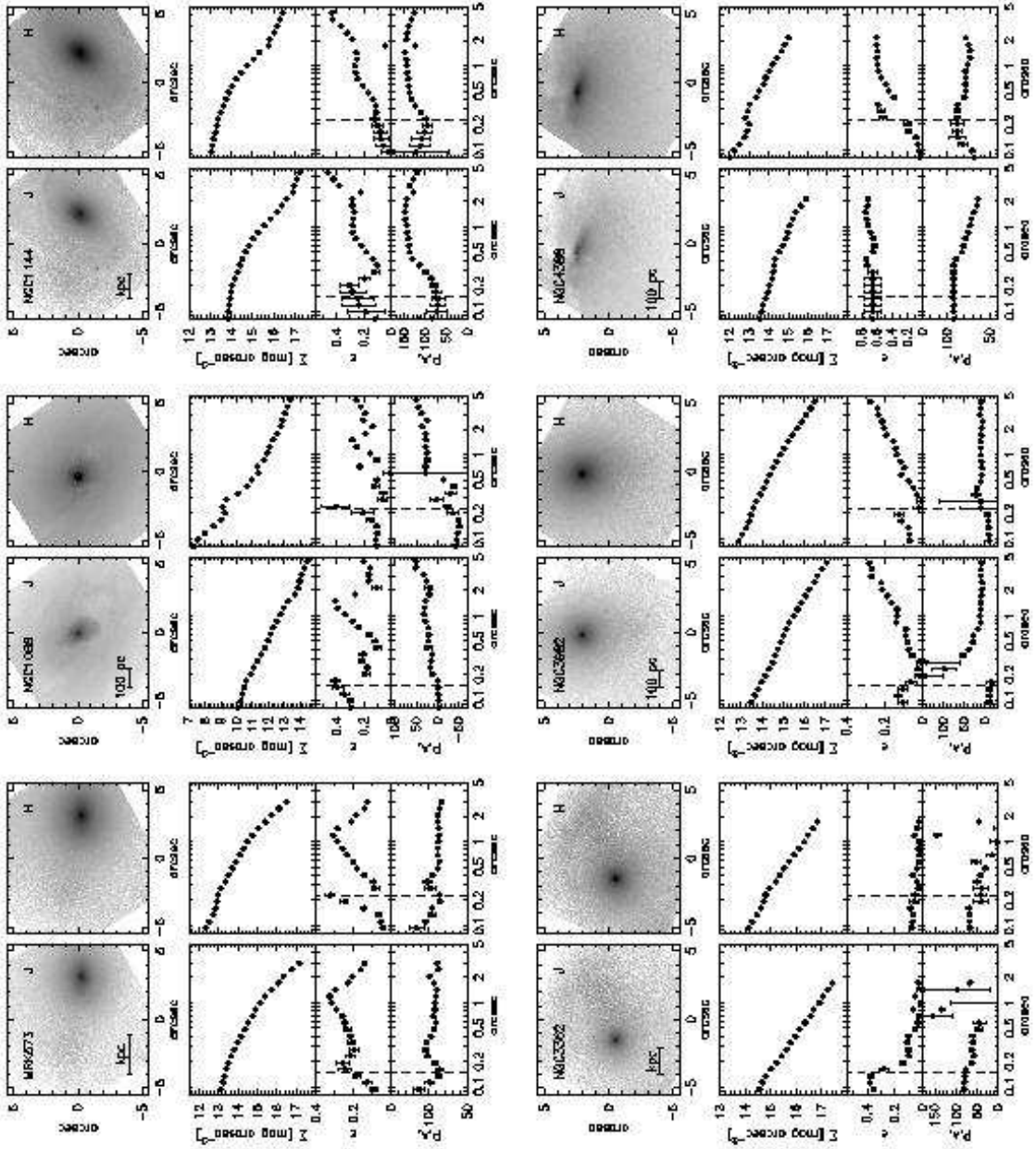}
\caption{Same as Figure~\ref{fig1} for Mrk~573, NGC~1068, NGC~1144, 
NGC~3362, NGC~3982, and NGC~4388. \label{fig3} }
\end{figure}

\begin{figure} 
\epsscale{1.}
\plotone{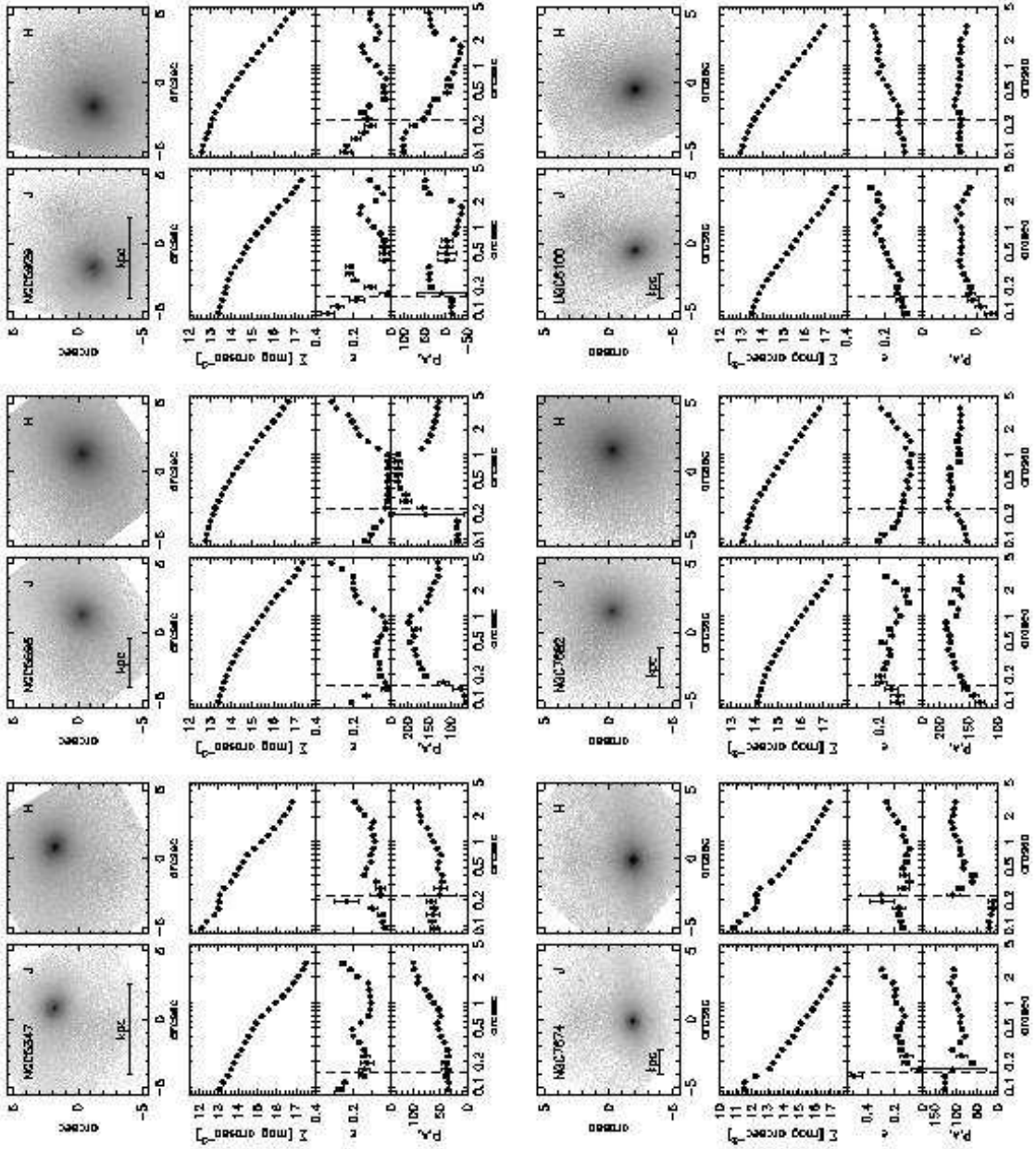}
\caption{Same as Figure~\ref{fig1} for NGC~5347, NGC~5695, NGC~5929, 
NGC~7674, NGC~7682, and UGC~6100. \label{fig4} }
\end{figure}

\begin{figure} 
\epsscale{1.}
\plotone{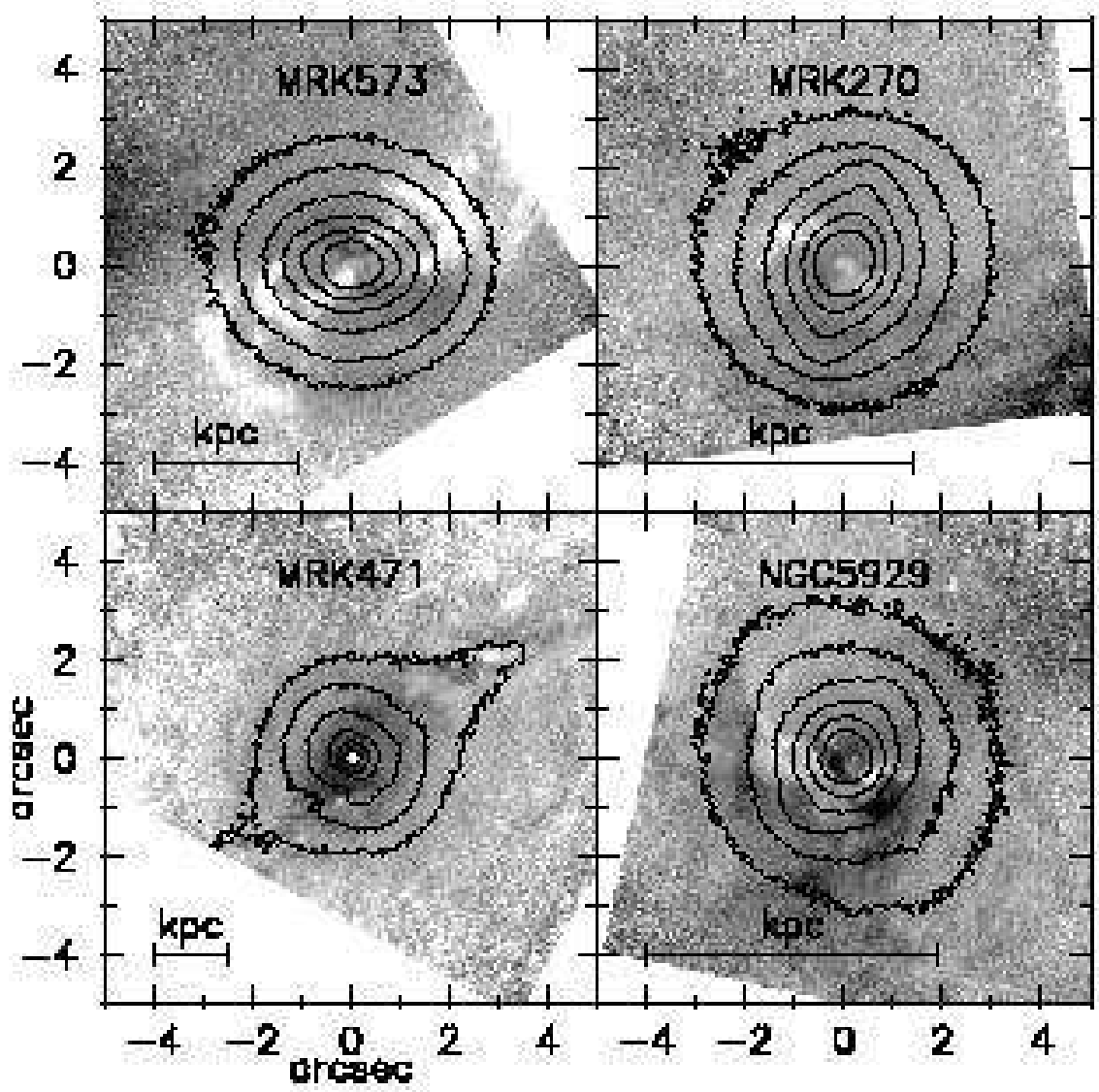}
\caption{$V-H$ colormaps and $H-$band surface brightness contours of the 
four nuclear bar candidates. In Mrk~573 the two spiral dust lanes (dark 
on this greyscale) appear to trace the leading edge of the nuclear bar. 
At larger radii these two dust lanes are lit up by the nuclear source and 
form the ionization cone. They turn blue on this colormap due to the 
presence of bright H$\alpha$ emission in the F606W filter bandpass.  
Mrk~270 and Mrk~471 show weaker evidence for an association between the 
dust morphology and the nuclear bar, while the dust morphology in NGC~5929 
exhibits no obvious, coherent structure. 
\label{fig5} }
\end{figure}

\noindent
Mrk 266  -- Mrk 266SW is in the midst of a merger and exhibits a very 
chaotic appearance which precludes a reliable elliptical isophote fit. 
The NIR peak shown in Figure~\ref{fig2} is actually offset from the peak in 
the visible-wavelength F606W image. The extremely irregular morphology and 
red nucleus of this 
galaxy strongly suggests that host galaxy material is obscuring the 
line of sight to the active nucleus. 

\noindent
Mrk 270   -- There is a good nuclear bar candidate visible in the $J$ and $H$ 
images that corresponds to the 
drop in the ellipticity 
at $\sim 1.9''$, corresponding to a projected physical semimajor axis length 
of $\sim 350$ pc. 
There are two dust lanes in the nuclear region that trace the edges of the 
nuclear bar and then turn abruptly to cross the nucleus as a straight 
dust lane. Outside $2''$ the position angle changes to 
match that of the host galaxy bar. 

\noindent
Mrk 573   -- This is the best example of a double-barred galaxy in the 
sample. The nuclear bar candidate previously noticed by \citet{pogge93} and 
\citet{capetti96} is readily apparent in the NIR surface brightness. 
The dust lane morphology is similar to Mrk~471, where the dust 
lanes trace the edge of the nuclear bar and then turn abruptly to cross the 
nucleus perpendicular to the bar position angle as a straight dust lane. 

\noindent
NGC 1068  -- The elliptical isophote fits to this galaxy were compromised 
by significant circumnuclear dust and the bright nuclear point source. The 
nuclear point source is particularly prominent at $H$, which supports the 
interpretation that this galaxy harbors a dust-obscured broad line region, 
first suggested by the polarization study by \citet{miller83}. While this 
nearly face-on galaxy shows significant radial ellipticity variations 
in $J$ and $H$, these variations are clearly uncorrelated as they reach 
a peak at 
$\sim 1.5''$ in $J$, but $\sim 0.7''$ in $H$. The much larger variation 
in $J$ is likely due to dust. The ellipticity variation at $0.7''$ in $H$ 
is coincident with a variation in the surface brightness profile and is due to 
the PSF diffraction pattern. 

\noindent
NGC 1144  -- This interacting system has relatively smooth NIR surface 
brightness distribution in its central few arcseconds, although the dust lane 
visible in the $J$ and $H$ grayscale images illustrates the more disturbed 
morphology at larger scales. 

\noindent
NGC 3362  -- Elliptical isophotes are a relatively poor fit to this galaxy 
outside of $\sim 1''$ as the galaxy is extremely faint. 

\noindent
NGC 3982  -- \citet{mcleod95} did not find evidence of the host galaxy bar 
in their $K$ image of this ``SAB'' galaxy. While our ellipse fits and 
GALFIT model do recover a larger ellipticity than that suggested by the 
axis ratio of $b/a \sim 0.9$ they measured, our images are not sufficiently 
sensitive on the large scales necessary to detect a host galaxy bar. 

\noindent
NGC 4388  -- This edge-on galaxy has an extremely chaotic appearance at 
small scales, at 
least partially due to a host galaxy dust lane passing to the immediate 
north of the nucleus and significantly attenuating even the $H-$band light. 
This galaxy cannot be well fit with elliptical isophotes. 

\noindent
NGC 5347  -- \citet{regan99} noticed a straight dust lane crossing the nucleus
in this galaxy and suggested that it contains a nuclear bar, although 
\citet{martini99} did not find this structure in their color maps. There 
is a slight jump in the ellipticity at $\sim 0.5''$, corresponding to a 
projected semimajor axis length of $\sim 80$ pc, but the strength of the 
ellipticity variation is much greater at $J$ than at $H$, suggesting that dust 
is responsible. Dust is clearly present to the immediate south of 
the nucleus in Figure~2 of \citet{martini99}, nearly perpendicular to the 
position angle of the nuclear bar and this dust is likely responsible for the 
distortion in the surface brightness distribution. This dust also corresponds 
to the curved dust lane observed by \citet{regan99} on which they base their 
identification of a {\it gaseous} nuclear bar.  

\noindent
NGC 5695  -- This Seyfert 2 is quite bright and well fit by a Nuker profile. 
The nuclear point source is sufficiently faint that GALFIT did not need 
to include a nuclear point source in the 2-D model. At large radii the 
ellipticity starts to increase at the position angle of the host galaxy bar.  

\noindent
NGC 5929  -- The nuclear bar candidate in this interacting system has a 
semimajor axis of $\sim 1.7''$, corresponding to a projected semimajor axis 
length of
$\sim 280$ pc. The ellipticity variation is stronger in the $J$ elliptical 
isophote fits than at $H$ and there is dust nearly perpendicular to the nuclear 
bar candidate in the \citet{martini99} color map. The inner region of this 
galaxy is sufficiently regular to be well-fit by GALFIT and the fit residuals 
show the signature of the nuclear bar. 

\noindent
NGC 7674  -- Polarized, broad emission lines were first detected in this 
object by \citet{miller90}. The extremely bright nuclear point source 
compared to the other Seyfert 2s (with the exception of NGC~1068) reinforce 
its interpretation as an obscured Seyfert~1. The elliptical isophotes are 
a reasonable match to the surface brightness distribution of this galaxy 
outside the immediate influence of the nuclear source. 

\noindent
NGC 7682  -- This Seyfert 2 was not well fit by a Nuker profile and instead 
was fit with a S\'ersic profile. The nuclear point source is sufficiently 
faint that GALFIT did not need to include it in the 2-D fit to the 
surface brightness distribution. 


{}

\clearpage

\clearpage

\clearpage

\clearpage

\clearpage

\clearpage
\begin{deluxetable}{lccccccc}
\tabletypesize{\scriptsize}
\tablenum{1}
\tablewidth{500pt}
\tablecaption{Sample Characteristics \label{tbl:sample} }
\tablehead {
  \colhead{Name} &
  \colhead{Other ID} &
  \colhead{Seyfert Type} &
  \colhead{Galaxy Type} &
  \colhead{Note} &
  \colhead{Distance} &
  \colhead{pc/$''$} \\
}
\startdata
Mrk 334  &       & 1.8   & Pec          & Disturbed     &  88.4 & 429 \\
Mrk 471  &       & 1.8   & SBa          &               & 137.3 & 666 \\
Mrk 744  &NGC 3786& 1.8  & SAB(rs)a pec & Interacting   &  36.1 & 175 \\
UGC 12138&2237+07  & 1.8 & SBa          &               & 102.8 & 498 \\
NGC 5033 &       & 1.9\tablenotemark{a}   & SA(s)c       & Edge On       &  21.3 & 103 \\
NGC 5252 &       & 1.9   & SO           &               &  90.7 & 440 \\
NGC 5273 &       & 1.9\tablenotemark{a}   & SA(s)0$^0$   &               &  16.5 &  80 \\
NGC 5674 &       & 1.9   & SABc         &               &  98.1 & 476 \\
UM 146   &0152+06& 1.9   & SA(rs)b      &               &  71.6 & 347 \\
Mrk 461  &       & 2     & S            &               &  65.6 & 318 \\
Mrk 266  &NGC5256& 2     & Comp Pec     &               & 111.0 & 538 \\
Mrk 270  &NGC5283& 2     & SO?          &               &  38.2 & 185 \\
Mrk 573  &       & 2     & (R)SAB(rs)O+:&               &  71.0 & 344 \\
NGC 1068 &       & 2     & SAb          &               &  14.4 &  70 \\
NGC 1144 &       & 2     & RingB        & Interacting   & 116.4 & 564 \\
NGC 3362 &       & 2     & SABc         &               & 108.7 & 527 \\
NGC 3982 &       & 2     & SAB(r)b:     &               &  17.0 &  82 \\
NGC 4388 &       & 2     & SA(s)b: sp   & Edge On       &  16.8 &  81 \\
NGC 5347 &       & 2     & (R')SB(rs)ab &               &  31.4 & 152 \\
NGC 5695 &Mrk 686& 2     & SBb          &               &  56.9 & 276 \\
NGC 5929 &       & 2     & Sab: pec     & Interacting   &  34.9 & 169 \\
NGC 7674 &Mrk 533& 2     & SA(r)bc pec  &               & 118.5 & 575 \\
NGC 7682 &       & 2     & SB(r)ab      &               &  70.8 & 343 \\
UGC 6100 &A1058+45& 2    & Sa?          &               & 117.6 & 570 \\
\enddata
\tablenotetext{a}{Classified as type 1.5 by \citet{ho97b}}
\tablecomments{
Properties of the galaxies observed with the NICMOS Camera 1 for program 
GO7867, along with NGC~1068.
Columns 1 \& 2 list the most common names for the targets, and column 3
lists its Seyfert type as reported by Osterbrock \& Martel (1993). In
column 4 we have compiled the morphological type for the host galaxy
from NED, while in column 5 we have provided additional comments on the 
galaxy morphology.  Column 6 lists the distance of the galaxy in Mpc 
\citep[from][or using \citet{yahil77}, assuming $H_0 = 75$ km s$^{-1}$ 
Mpc$^{-1}$]{tully88,ho97c,tonry01}.
Column 7 contains the projected size in parsecs of one arcsecond at the 
distance of the galaxy.
}
\end{deluxetable}
\normalsize

\begin{center}
\begin{deluxetable}{lcccc}
\tabletypesize{\scriptsize}
\tablenum{2}
\tablewidth{500pt}
\tablecaption{Properties of the Nuclear Bar Candidates\label{tbl:bars}}
\tablehead {
  \colhead{Name} &
  \colhead{PA [degree]} &
  \colhead{$a$ [arcsec]} &
  \colhead{$a_p$ [parsec]} &
  \colhead{Host bar PA} \\
}
\startdata
Mrk 573   . . . . . . . . . . & 90  & 1.2 & 410 &   0  \\
Mrk 270   . . . . . . . . . . & 160 & 1.9 & 350 &      \\
NGC 5929  . . . . . . . . . .& 150 & 1.7 & 280 &      \\
Mrk 471   . . . . . . . . . . & 60  & 1.3 & 860 & 130  \\
\enddata
\tablecomments{
Properties of the nuclear bars in our sample. Column 2 lists the position 
angle (north through east) of the nuclear bar candidates for each of the 
galaxies in column 1. Columns 3 and 4 provide the semimajor axis length of the 
bar in arcseconds and the corresponding projected size in parsecs. 
Column 5 gives the position angle of the host galaxy bar, if present. 
} 
\end{deluxetable}
\normalsize
\end{center}

\begin{deluxetable}{lccccccc}
\tabletypesize{\scriptsize}
\tablenum{3}
\tablewidth{500pt}
\tablecaption{Galaxy Profile Fits\label{tbl:profs} }
\tablehead {
  \colhead{Name} &
  \colhead{$m_H^{nuc}$ [mag]} &
  \colhead{$\mu_b$ [mag/arcsec$^2$]} &
  \colhead{$r_b$ ($r_s$) [$''$]} &
  \colhead{$\alpha$ ($n$)} &
  \colhead{$\beta$} &
  \colhead{$\gamma$} &
  \colhead{Function} \\
}
\startdata
Mrk 270	 & 15.8 &      & 0.61 & 1.74 &      &      & S\'ersic \\
Mrk 573  & 15.9 & 13.9 & 0.61 & 2.95 & 1.54 & 0.70 & Nuker \\
NGC 3982 & 16.6 & 14.5 & 0.73 & 4.96 & 1.08 & 0.71 & Nuker \\
NGC 5252 & 15.8 & 15.6 & 2.12 & 0.28 & 1.71 & 0.52 & Nuker \\
NGC 5273 & 14.9 & 14.2 & 0.80 & 9.18 & 1.37 & 0.60 & Nuker \\
NGC 5674 & 13.9 & 13.9 & 0.45 & 4.41 & 1.50 & 0.46 & Nuker \\
NGC 5695 &      & 13.8 & 0.45 & 3.55 & 1.38 & 0.66 & Nuker \\
NGC 5929 & 17.1 & 13.3 & 0.29 & 1.96 & 1.40 & 0.46 & Nuker \\
NGC 7674 & 13.4 & 16.7 & 2.60 & 0.18 & 2.23 & 0.45 & Nuker \\
NGC 7682 &      &      & 0.41 & 1.94 &      &      & S\'ersic \\
UGC 6100 &      & 12.7 & 0.11 & 1.88 & 1.45 & 0.49 & Nuker \\
\enddata
\tablecomments{
Galaxy profile fits to the F160W images. For each galaxy in column 1 we 
list the $H$ magnitude of the nuclear PSF in column 2 and the best fitting 
parameters of a Nuker (S\'ersic) profile in columns 3 -- 7. Column 
8 identifies which function was fit to each galaxy. These fitting functions 
are defined in section~\ref{sec:analysis}. 
}
\end{deluxetable}
\normalsize

\end{document}